\documentclass[twocolumn,11pt]{article}
\usepackage{amssymb,amsmath,latexsym}

\usepackage[square,numbers]{natbib}
\usepackage{chemarr}
\usepackage{enumitem}
\usepackage{graphicx}
\usepackage{amssymb}
\usepackage{lineno}
\usepackage{amsmath}
\usepackage{sidecap}
\usepackage{wrapfig}
\usepackage{multirow}
\usepackage{float}
\begin{document}
\date{\today}
\newcommand{\MSun}{{M_\odot}}
\newcommand{\LSun}{{L_\odot}}
\newcommand{\Rstar}{{R_\star}}
\newcommand{\calE}{{\cal{E}}}
\newcommand{\calM}{{\cal{M}}}
\newcommand{\calV}{{\cal{V}}}
\newcommand{\calO}{{\cal{O}}}
\newcommand{\calH}{{\cal{H}}}
\newcommand{\calD}{{\cal{D}}}
\newcommand{\calB}{{\cal{B}}}
\newcommand{\calK}{{\cal{K}}}
\newcommand{\labeln}[1]{\label{#1}}
\newcommand{\Lsolar}{L$_{\odot}$}
\newcommand{\farcmin}{\hbox{$.\mkern-4mu^\prime$}}
\newcommand{\farcsec}{\hbox{$.\!\!^{\prime\prime}$}}
\newcommand{\kms}{\rm km\,s^{-1}}
\newcommand{\cc}{\rm cm^{-3}}
\newcommand{\Alfven}{$\rm Alfv\acute{e}n$}
\newcommand{\Vap}{V^\mathrm{P}_\mathrm{A}}
\newcommand{\Vat}{V^\mathrm{T}_\mathrm{A}}
\newcommand{\D}{\partial}
\newcommand{\DD}{\frac}
\newcommand{\TAW}{\tiny{\rm TAW}}
\newcommand{\mm }{\mathrm}
\newcommand{\Bp }{B_\mathrm{p}}
\newcommand{\Bpr }{B_\mathrm{r}}
\newcommand{\Bpz }{B_\mathrm{\theta}}
\newcommand{\Bt }{B_\mathrm{T}}
\newcommand{\Vp }{V_\mathrm{p}}
\newcommand{\Vpr }{V_\mathrm{r}}
\newcommand{\Vpz }{V_\mathrm{\theta}}
\newcommand{\Vt }{V_\mathrm{\varphi}}
\newcommand{\Ti }{T_\mathrm{i}}
\newcommand{\Te }{T_\mathrm{e}}
\newcommand{\rtr }{r_\mathrm{tr}}
\newcommand{\rbl }{r_\mathrm{BL}}
\newcommand{\rtrun }{r_\mathrm{trun}}
\newcommand{\thet }{\theta}
\newcommand{\thetd }{\theta_\mathrm{d}}
\newcommand{\thd }{\theta_d}
\newcommand{\thw }{\theta_W}
\newcommand{\beq}{\begin{equation}}
\newcommand{\eeq}{\end{equation}}
\newcommand{\bsplit}{\begin{split}}
\newcommand{\esplit}{\end{split}}
\newcommand{\ben}{\begin{enumerate}}
\newcommand{\een}{\end{enumerate}}
\newcommand{\bit}{\begin{itemize}}
\newcommand{\eit}{\end{itemize}}
\newcommand{\barr}{\begin{array}}
\newcommand{\earr}{\end{array}}
\newcommand{\bc}{\begin{center}}
\newcommand{\ec}{\end{center}}
\newcommand{\DroII}{\overline{\overline{\rm D}}}
\newcommand{\DroI}{{\overline{\rm D}}}
\newcommand{\eps}{\epsilon}
\newcommand{\vepsdi}{{\cal E}^\mathrm{d}_\mathrm{i}}
\newcommand{\vepsde}{{\cal E}^\mathrm{d}_\mathrm{e}}
\newcommand{\lraS}{\longmapsto}
\newcommand{\lra}{\longrightarrow}
\newcommand{\LRA}{\Longrightarrow}
\newcommand{\Equival}{\Longleftrightarrow}
\newcommand{\DRA}{\Downarrow}
\newcommand{\LLRA}{\Longleftrightarrow}
\newcommand{\diver}{\mbox{\,div}}
\newcommand{\grad}{\mbox{\,grad}}
\newcommand{\cd}{\!\cdot\!}
\newcommand{\Msun}{{\,{\cal M}_{\odot}}}
\newcommand{\Mstar}{{\,{\cal M}_{\star}}}
\newcommand{\Mdot}{{\,\dot{\cal M}}}
\newcommand{\ds}{ds}
\newcommand{\dt}{dt}
\newcommand{\dx}{dx}
\newcommand{\dr}{dr}
\newcommand{\dth}{d\theta}
\newcommand{\dphi}{d\phi}

\newcommand{\pt}{\frac{\partial}{\partial t}}
\newcommand{\pk}{\frac{\partial}{\partial x^k}}
\newcommand{\pj}{\frac{\partial}{\partial x^j}}
\newcommand{\pmu}{\frac{\partial}{\partial x^\mu}}
\newcommand{\pr}{\frac{\partial}{\partial r}}
\newcommand{\pth}{\frac{\partial}{\partial \theta}}
\newcommand{\pR}{\frac{\partial}{\partial R}}
\newcommand{\pZ}{\frac{\partial}{\partial Z}}
\newcommand{\pphi}{\frac{\partial}{\partial \phi}}

\newcommand{\vadve}{v^k-\frac{1}{\alpha}\beta^k}
\newcommand{\vadv}{v_{Adv}^k}
\newcommand{\dv}{\sqrt{-g}}
\newcommand{\fdv}{\frac{1}{\dv}}
\newcommand{\dvr}{{\tilde{\rho}}^2\sin\theta}
\newcommand{\dvt}{{\tilde{\rho}}\sin\theta}
\newcommand{\dvrss}{r^2\sin\theta}
\newcommand{\dvtss}{r\sin\theta}
\newcommand{\dd}{\sqrt{\gamma}}
\newcommand{\ddw}{\tilde{\rho}^2\sin\theta}
\newcommand{\mbh}{M_{BH}}
\newcommand{\dualf}{\!\!\!\!\left.\right.^\ast\!\! F}
\newcommand{\cdt}{\frac{1}{\dv}\pt}
\newcommand{\cdr}{\frac{1}{\dv}\pr}
\newcommand{\cdth}{\frac{1}{\dv}\pth}
\newcommand{\cdk}{\frac{1}{\dv}\pk}
\newcommand{\cdj}{\frac{1}{\dv}\pj}
\newcommand{\rad}{\;r\! a\! d\;}
\newcommand{\half}{\frac{1}{2}}
\newcommand{\ARZL}{\textquotedblleft}
\newcommand{\ARZR}{\textquotedblright}
\twocolumn[
  \begin{@twocolumnfalse}
  \title{Why the spacetime embedding incompressible supradense nuclear matter  is conformally flat?}

\author{\thanks{E-mail:AHujeirat@uni-hd.de}~Hujeirat  A.A., \thanks{ravi.samtaney@kaust.edu.sa} Samtaney, R. \\
IWR, Universit\"at Heidelberg, 69120 Heidelberg, Germany \\
Applied Mathematics and Computational Science, CEMSE Division, KAUST, Saudi Arabia}
\maketitle
\begin{abstract}
\noindent The multi-messenger observations of the  merger event in GW170817  did not rule out the possibility that the remnant might be
a dynamically stable neutron star with $\mathcal{M}_{rm}\geq 2.69  \mathcal{M}_{\odot}.$  Based on this and other recent events,
 I argue that the universal maximum density hypothesis should be revived.  Accordingly,
the central densities in the cores of ultra-compact objects must be upper-limited  by the critical density number  $n_{cr},$ beyond which supradense nuclear
 matter becomes purely incompressible.\\
Based on the spacetime-matter coupling  in GR, it is shown that the topology of spacetime embedding  incompressible quantum fluids with
$n =n_{cr}$ must be Minkowski flat, which implies that spacetime at the background of ultra-compact objects should be of bimetric type.
\begin{center}
\textbf{Keywords:}{~~Relativity: numerical, general, black hole physics --pulsars-- neutron stars--pulsars--- superfluidity --superconductivity--incompressibility--gluons--quarks--plasmas--- QCD }
\end{center}
\end{abstract}
 \end{@twocolumnfalse}
 ]

\textbf{Most} theoretical investigations indicate that pulsars, neutron stars and magnetars, that comprise the family of ultra-compact objects (UCOs),
whose masses $\mathcal{M}_{NS}\geq 1.4 \MSun,$ should have central densities much larger than the nuclear one [\citealp{Ozel2016}-\nocite{Ozel2016,Camenzind2007,HujeiratMassiveNSs18,Hujeiratmetamorph2018,Raithel2019}\citealp{Raithel2019}].
 Theoretically this posses an upper limit of their
  maximum mass $\mathcal{M}^{NS}_{max} \leq 2.3 \MSun$ for almost all  EOSs, though non of these
  objects have been ever observed with  $\mathcal{M}^{NS}_{max} \geq 2.1 \MSun$ (see \cite{Ozel2016} and the references therein).

   Indeed, the multi-messenger observations of the  merger of the two neutrons stars in GW170817  did not rule the formation of
  a massive NS with $\mathcal{M}^{NS}\approx  2.79 \MSun$
  [\citealp{Raithel2019}-\nocite{PiroXray2019,AbbottNSProberties2019}\citealp{HujSam2020B}], hence the mechanisms that limit their theoretical mass  range must be  revisited.\\
Let us assume that there exist a universal maximum energy density $\varepsilon_{max}$, beyond which supradense nuclear
fluids becomes incompressible. Under these circumstances the solution of the field equations:
\beq
R^{\mu\nu} - \DD{1}{2} g^{\mu\nu} R = 8 \pi G\, T^{\mu\nu}
\eeq
reads:
\beq
ds^2 =  g_{\mu\nu}dx^\mu dx^\nu = \DD{1}{S^\kappa}  dt^2   -    \DD{dr^2}{S}  - r^2 d\Omega^2,
\eeq
where $S = 1-r_s/r,~\kappa = \DD{1+3\alpha_0}{2},\ \alpha_0 = P^L_b/ \varepsilon_{max}$ and $ \Omega^2= d\theta^2 + sin^2(\varphi).$
$P^L_b$ here is the local baryonic pressure.\\
In order to find out whether this spacetime is Weyl-transformable into a conformally flat spacetime of the type:
\beq
ds^2 =  \hat{g}_{\mu\nu}d\hat{x}^\mu d\hat{x}^\nu =  e^f ( d\tau^2   -    d\rho^2 - \rho^2 d \Sigma^2),\\
\eeq

we need to analyse the relation between both coordinate systems, solve for the conformal factor  $e^f = F(\tau, \rho,\vartheta, \psi)$
and  discuss their physical consistences (see \cite{EmreDil2016,Ramos2019} and the references therein). As  the line element, $ds,$ should be invariant under the transformation, both metrics are related to each other through:
\beq
                           \hat{g}_{\mu\nu} = \DD{\D x^\mu}{\D \hat{x}^\nu}   \DD{\D x^\mu}{\D \hat{x}^\nu}    g_{\mu\nu}.
 \eeq
 Using  algbraic manipulation and setting $ \DD{\D x^\mu}{\D \hat{x}^\nu} = \delta^\mu_\nu,$ the following equalities are obtained:
 \beq
 e^f =   (\DD{1}{S})^\kappa  (\DD{\D t}{\D \tau} )^2  =   (\DD{1}{S}) (\DD{\D r}{\D \varrho} )^2 =    ( \DD{r}{\rho})^2   (\DD{\D \Omega}{\D \Sigma} )^2.
 \eeq
A strictly flat spacetime would correspond to $e^f=1.$ In this case, the relation between both coordinates systems read:
  \beq
  \barr{lllll}
    \DD{d \tau}{d t}  &= & \DD{1}{S^{\kappa/2 }} & \Leftrightarrow &   d \tau  = \DD{1}{(1-a_0 r^2)^{\kappa/2 }}dt \\
  \DD{d \rho}{d r} &= & \DD{1}{\sqrt{S}}            &  \Leftrightarrow  &  d \rho  = \DD{1}{\sqrt{1-a_0 r^2}} dr \\
                              &  &                                             & \Rightarrow   &        \rho   =  \DD{ arcsin (\sqrt{a_0} r)} {\sqrt{a_0} }r \\
  \DD{d \Sigma}{d \Omega} & = & (\DD{1}{S})    & \Leftrightarrow &   d \Sigma  = \DD{\sqrt{a_0} r}{ arcsin (\sqrt{a_0} r)} \,d\Omega \\
  \earr
  \eeq
   Consequently, the spacetime:
  \beq
ds^2 =  (\DD{1}{S})^{\kappa/2}  dt^2   -     (\DD{1}{S})dr^2  - r^2 d\Omega^2,
\eeq
appears to be conformal to the flat spacetime:
\beq
ds^2 =  d\tau^2   -    d\rho^2 - \rho^2 d \Sigma^2.
\eeq
Mathematically, while $\kappa$ may accept  other values, but  $\kappa= 1 $ is most reasonable as the volume-expansion rate of the incompressible core in both dual-spacetimes is equal and independent of the core's compactness, i.e.,
\beq
 \DD{d \rho}{d\tau} = \DD{d r}{d t}.
\eeq
In most cases, modelings of  the interiors of UCOs usually relay on using one single EOS throughout the entire object and obeying: $P^L_b \leq 1/3\, \varepsilon.$
This requires a strong spatial variation  of $\varepsilon$  in the neighborhood of $r=0,$ where the regularity condition is imposed.
However this egularity condition is practically equivalent  to imposing zero-compressibility., i.e. $\nabla P = \nabla \varepsilon|_{r=0} =0$. To overcome this inconsistency, the central density is manually
increased to be much higher than the nuclear number density, $n_0,$ therefore giving rise  to causality violation.
On the other hand, recently it was suggested that at $n_{cr}\approx 3\times n_0$ and zero-temperature,
the neutrons at the very central regions of massive pulsars  ought to undergo a phase transition, through which they merge together to form embryonic super-baryons (:SBs,  see Fig. 1 as well as \cite{HujeiratMassiveNSs18,Hujeiratmetamorph2018} and the references therein).  The number density of the enclosed incompressible gluon-quark superfluid would attain the value   $n_{cr},$ which could be easily  reached by NSs with moderate masses.
The state of the fluid on the verge of  merger is said to be maximally compressible (MC) and the corresponding stress-energy tensor becomes traceless: $\textbf{T} = \varepsilon_{cr} - 3 P^L_b = 0.$
Although the MC state  is expected to be a short-living transient phase,   $\textbf{T}=0$ insures that the matter-field correspondence, and specifically the
incompressibility character of the quantum fluid,  is invariant under Weyl and conformal transformations.\\
\begin{figure}[t]
\centering {\hspace*{-0.15cm}
\includegraphics*[angle=-0, width=7.0cm]{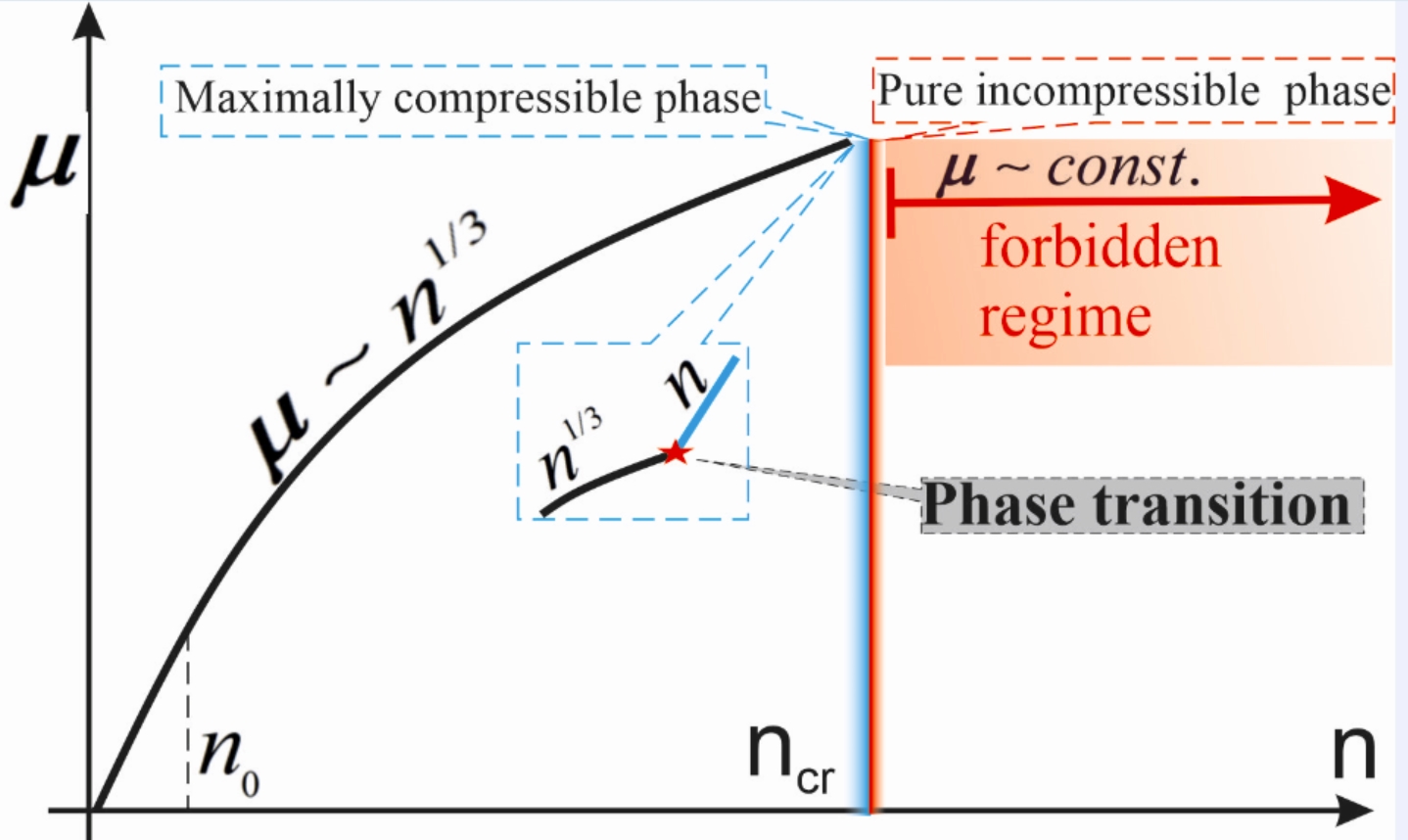}
}
\caption{\small The development of the chemical potential $\mu$ of ultradense quantum fluids with the number density $n.$  For  $n_0 \leq n < _{cr},$ $\mu \sim n^{1/3}.$
As $n \rightarrow n_{cr} $ the quantum fluid converges to the maximally compressible state and undergoes a phase transition into
the incompressible state, where $P=\varepsilon_{cr}= c_0 n_{cr}^2.$ The forbidden region correspond to number densities that are  beyond
the universal maximum allowable value $n_{cr}.$
}\label{NSInternal}
\end{figure}
 When the compressible matter surrounding the cores of NSs cools down on the cosmic time,  the curved spacetime at the background would
compress the baryons at the center together, thereby decreasing  the separation distance between the
 baryons, $d_b,$ down to values  comparable to the average distance $d_g,$ between quarks (see Fig. 2).
In this case, a free dark energy $\Delta \varepsilon_b^+$ of order   $ c \hbar / 2 d_b $ is generated.  For $d_b = d_q, $  $\Delta \varepsilon_b^+$ is comparable to the rest energy of an isolated baryon, i.e.,
\beq
   \DD{\Delta \varepsilon_b^+}{\Delta \varepsilon_q } \sim \DD{2 d_q}{d_q + d_b} \rightarrow
    \left\{ \begin{array}{ll} 0\,;   &  d_b \gg d_q \\
                                           1\,;  &  d_b = d_q,\\
\end{array} \right.
\eeq
where $\Delta \varepsilon_q   = 0.939$ GeV.  Hence $\Delta \varepsilon_b^+$ is capable of deconfining  the quarks inside baryons, thereby rendering merger
of the baryons possible.  Moreover, the pressure $P_{UP},$ induced by the uncertainty principle,  and  $\Delta \varepsilon_b^+$ become  duals
that  oppose compression/contraction of the core's matter, whilst enhancing the effective mass of the core.
The state of matter in the post merger phase is governed by EOS: $P_{UP}=  \Delta \varepsilon_b^+ = \varepsilon_{cr} = a_0 n^2_{cr}.$
The mass of the field quanta goes to zero and therefore
the quarks  communicate with each other via the massless gluons  at the speed of light  \cite{Zeldowich1962, Bludman1968}.

The dark energy contribution  $\triangle \varepsilon^+$ must be stored locally; it goes specifically into enhancing the surface tension of the super-baryon
and acts as a confining force for  the enclosed ocean of quarks. At a certain point of the cosmic time,  the core would decay and dark
energy would be liberated,  thereby provoking  a  hadronization process of the guon-quark superfluid .\\
\begin{figure}[t]
\centering {\hspace*{-0.15cm}
\includegraphics*[angle=-0, width=7.7cm]{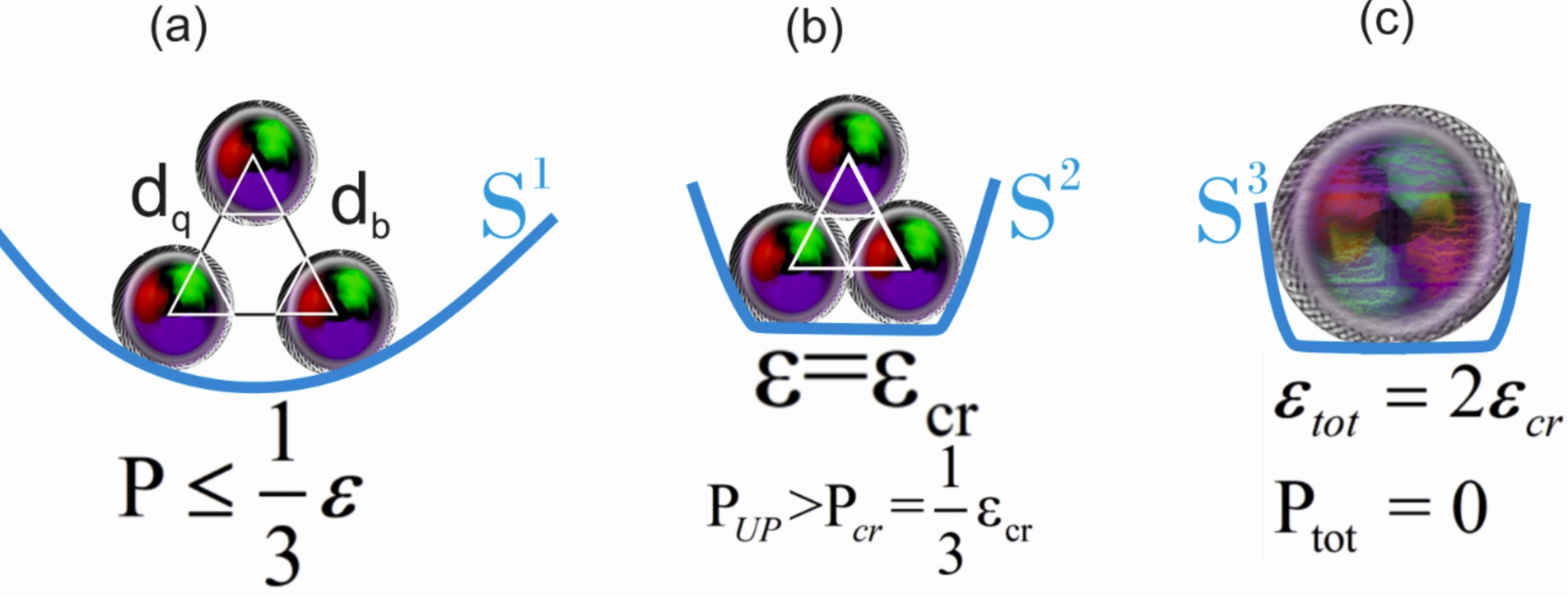}
}
\caption{\small A schematic description of the spacetime curvature (denoted with S) as opposed to the EOS and number density.
$d_b$ and $d_q$ denote the mean distances between two arbitrary baryons and quarks, respectively. The case (a) corresponds to
compressible quantum fluid embedded in a Schwarzschild-like spacetime $(S^1)$, (b) to the maximally compressible fluid phase, where the
boundaries of baryons overlap, i.e. $d_b = d_g$ and the pressure $P_{UP}$ induced  by the uncertainty principle becomes comparable
to $\varepsilon_{cr}$. Here the spacetime start flattening  to become nearly a Minkowski-type spacetime $(S^2)$. (c) corresponds to the final state in which baryons start merging to form a super-baryon, whose interior is made of incompressible gluon-quark superfluid embedded
by a strictly flat spacetime $(S^3)$.
}\label{NSInternal}
\end{figure}
For a given  super-baryon resulting from  the merger of N-baryons,  the effective energy is expected  to be:
 \beq
 {\Large \varepsilon}^{tot}_{\tiny{SB}} \approx  2\,  N\times \varepsilon_0.
 \eeq
This appears to be in line with the short-living pentaquark formation observed at the LHC-experiment  (see \cite{LHCb2015} and the references therein), though  the entropy and density regimes are totally different from those in the cores of UCOs.\\
Noting that the core must be a  3D spherically symmetric with zero-entropy enclosed matter which behaves as a single quantum entity embedded in a flat spacetime,  the governing physics is predicated to be mirrored onto its two-dimensional surface  in accord with  the holographic principle.
However, it is not clear at all, how and what kind of
 information could be still storable  on the surface under zero-entropy conditions?\\

Finally,  when combining the result-presented  here with the following arguments:
\bit
\item The remnant of GW170817 didn't necessarily collapse into a BH
\item BHs with $\mathcal{M}_{BH}\leq 5 \MSun$ may  safely ruled out
\item The first generation of stars may have formed massive pulsars that should be dark by now
\item  The glitch phenomena in pulsars are triggered by  topologal changes the of bimetric spacetime
          embedding pulsars  (see \cite{HujSam2020C} and the references therein),
\eit
 then the existence of a universal maximum energy density is an inevitable  conclusion.\\

It should be noted  however, that the existence of a universal maximum density, $n_{cr},$ does not necessary rule out BHs as astrophysical
objects, whose existence is observationally well-verified, however, it argues against their classical formation scenario
as well as against the existence of  matter singularities at their centers.\\

 Dr. Dil, A. is gratefully acknowledged for valuable discussions.

 \end{document}